\input phyzzx.tex
\tolerance=1000
\voffset=-0.0cm
\hoffset=0.7cm
\sequentialequations
\def\rl{\rightline}

\def\t1{{\tilde 1}}

\REF{\DIN}{E. Copeland, A. Liddle, D. Lyth, E. Stewart and D. Wands, {\it Phys.
Rev}
{\bf D49} (1994) 6410; M. Dine, L. Randall and S. Thomas, {\it
Phys. Rev . Lett.} {\bf 75} (1995) 398.}
\REF{\HBD}{E. Halyo, {\it Phys. Lett.} {\bf B387}  (1996) 43, hep-ph/9606423;
P. Binetruy and G. Dvali, {\it Phys. Lett.} {\bf B388}  (1996) 241,
hep-ph/9606342.}
\REF{\CAS}{J. A. Casas and C. Munoz, {\it Phys. Lett.} {\bf B216} (1989) 37; J.
A. Casas, J. Moreno, C. Munoz and M. Quiros, {\it Nucl. Phys}
{\bf B328} (1989) 272.}
\REF{\DSW} {M. Dine, N. Seiberg and E. Witten, {\it Nucl. Phys} {\bf B289}
(1987) 585.}
\REF{\HI}{A. D. Linde, {\it Phys. Lett.} {\bf B259} (1991) 38; {\it Phys. Rev}
{\bf D49} (1994) 748.}
\REF{\ADS}{J. Atick, L. Dixon and A. Sen, {\it Nucl. Phys} {\bf B292} (1987)
109.}
\REF{\KMR}{C. Kolda and J. March--Russell, hep-ph/9802358.}
\REF{\KL}{C. Kolda and D. Lyth, hep-ph/9812234.}
\REF{\RIO}{J. Espinoza, A. Riotto and G. Ross, {\it Nucl. Phys.} {\bf B531}
(1998) 461;
hep-ph9804214.}
\REF{\LR}{D. Lyth and A. Riotto, hep-ph/9807278.}
\REF{\AFIV}{G. Aldazabal, A. Font, L. Ibanez and G. Violero, hep-th/9804026.}
\REF{\IMR}{L. Ibanez,  C. Munoz and S. Rigolin, hep-ph/9812397.}
\REF{\IRU}{L. Ibanez, R. Rabadan and A. Uranga, hep-th/9808139.}
\REF{\JMR}{J. March--Russell, {\it Phys. Lett.} {\bf B437}  (1998) 318,
hep-ph/9806426.}

\singlespace
\rl{SU-ITP-99-2}
\rl{hep-ph/9901302}
\rl{\today}
\pagenumber=0
\normalspace
\medskip
\bigskip
\titlestyle{\bf{ D--term Inflation in Type I String Theory}}
\smallskip
\author{ Edi Halyo{\footnote*{e--mail address: halyo@dormouse.stanford.edu}}}
\smallskip
\centerline {Department of Physics}
\centerline{Stanford University}
\centerline {Stanford, CA 94305}
\smallskip
\vskip 2 cm
\titlestyle{\bf ABSTRACT}

D-term inflation realized in heterotic string theory has two problems: the
scale of the
anomalous D-term is too large for accounting for  COBE data and the coupling
constant of the anomalous $U(1)$ is too large for supergravity to be valid. We
show that
both of these problems can be easily solved in D-term inflation based on type I
string theory or orientifolds of type IIB strings.

\singlespace
\vskip 0.5cm
\endpage
\normalspace

\centerline{\bf 1. Introduction}
\medskip

Early attempts to incorporate inflationary cosmology in  string theory (or
supergravity) were
not very successful due to the inflaton mass problem, i.e. the mass of the
inflaton is generically as large as the Hubble constant when the vacuum energy
arises from F--terms[\DIN]. In this case, inflation cannot take place since the
slow--roll condition is violated.

An elegant solution to the inflaton mass problem in string theory (or
supergravity) is
D-term inflation[\HBD].\foot{For earlier work on D--term inflation see [\CAS].}
 In this scenario, the vacuum energy needed for inflation is dominated by
D-terms rather than F-terms. Thus, the inflaton mass problem is trivially
solved since the
dangerous contribution to the inflaton mass due to the F--terms is vanishing
(or negligible).
This can be easily realized in heterotic string theories since generically
there is an anomalous D-term arising from an anomalous $U(1)_A$ which can
contribute to the vacuum energy[\DSW]. In this scenario, the inflaton $\sigma$
is neutral under the anomalous $U(1)_A$ but has tree level couplings to other
fields $\phi, \bar \phi$
in the superpotential, i.e. $W=\lambda \sigma \phi \bar \phi$. The fields
$\phi, \bar \phi$ have $\pm1$ charges under
$U(1)_A$ and behave as the trigger fields in hybrid inflation models[\HI]. The
scalar potential
including the anomalous D-term is[\HBD]
$$V=|\lambda \sigma|^2(|\phi|^2+|\bar \phi|^2)+|\lambda \phi \bar \phi|^2+{g^2
\over 2}(|\phi|^2-
|\bar \phi|^2+M^2) \eqno(1)$$
where $\lambda \sim O(1)$ is a Yukawa coupling, $g$ is the gauge coupling of
$U(1)_A$ and
$M$ is the scale of the anomalous D-term. For heterotic strings it is given
by[\ADS]
$$M^2={1 \over {192 \pi^2}} g^2 (TrQ_A) M_P^2 \eqno(2)$$
Here $M_P \sim 2 \times 10^{18} ~GeV$ is the reduced Planck scale, $g \sim 1/2$
from gauge coupling unification and the trace is over the whole massless
spectrum of the string theory giving generically $TrQ_A \sim 100$.
Hybrid inflation occurs for large values of $\sigma$, $\sigma \sim M<<M_P$
which gives a
positive mass squared to $\phi, \bar \phi$ and forces them to have vanishing
VEVs. Then there
is a constant nonzero
vacuum energy  $V_0= g^2 M^2/2$ resulting in a period of inflation.
Supersymmetry is broken
by $V_0$ and as a result a one--loop potential for $\sigma$ is generated[\HBD]
$$V_{one-loop}(\sigma)={1 \over 2} g^2M^4 \left[1+{g^2 \over {8 \pi^2}} log
\left(\lambda \sigma
\over \Lambda \right) \right] \eqno(3)$$
above $\Lambda$ is the renormalization scale which does not affect the physics.
Due to this
potential the inflaton rolls slowly to its minimum during inflation. There is a
critical value
$\sigma_{cr}=gM/\lambda$ after which the mass squared of  $\bar \phi$ becomes
negative and it  rapidly falls to its new minimum at $\bar \phi=M$. This ends
inflation and restores supersymmetry.

It has been noted that the above D-term inflation scenario has two problems
when it is realized in
heterotic string theory[\KMR,\KL]\foot{For D-term inflation in explicit string
models see [\CAS,\RIO].}. The first problem arises from the magnitude of
density fluctuations
obtained from COBE data which requires[\KL,\LR]
$$(V_0/\epsilon)=6.7 \times 10^{16}~GeV \eqno(4)$$
where $\epsilon$ is one of the slow--roll parameters of inflation; $\epsilon={1
\over2} M_P^2 (V^{\prime}/V)^2$. For inflation with $N$ ($\sim 60$) e-folds one
needs
$$M \sim 8.5 \times 10^{15}GeV \times \left({50 \over N}\right)^{1/4}
\eqno(5)$$
However in heterotic string models with an anomalous D-term eq. (2) above
gives a scale too large to account for COBE data.

The second problem arises  from the fact that inflation can come to an end
before the inflaton
reaches its critical value if the second slow--roll parameter $\eta=M_P^2
|V^{\prime\prime}/V|^2$
becomes of order one (since slow--roll requires $\eta<<1$). This means
that[\KL,\LR]
$$\eta(\sigma)=\sqrt{\alpha \over {2 \pi}} {M_P \over \sigma} \eqno(6)$$
where $\alpha=g^2/4 \pi$. We see that $\eta \sim 1$ when
$$\sigma_f \sim \sqrt{\alpha \over {2 \pi}} M_P \eqno(7)$$
This gives $\sigma_f \sim M_P/10$ for the final value of $\sigma$ at the end of
inflation
which is much larger than $\sigma_{cr}$. Moreover, it can be shown that the
initial value
of the inflaton $\sigma_i$ needs to be
$$\sigma_i \sim \sqrt{\alpha N \over \pi} M_P \eqno(8)$$
which gives $\sigma_i \sim 0.8 M_P$. This is problematic because for Planckian
values of the inflaton one cannot use the effective low--energy supergravity
approximation. Actually,
this problem is probably less severe than it seems because the criterion should
not be $\sigma<M_P$ but rather that the one--loop corrections should be smaller
than
the tree level results (when the D--term is the source of vacuum energy and the
corrections
to the Kahler potential are not relevant). For example,
for $V_{one-loop}$ we see that even for $\lambda \sim 1$, $\sigma$ can be much
larger than
$M_P$ due to the factor $g^2/8 \pi^2$.\foot{We thank Andrei Linde for
clarifying this point.}

In this letter, we show that both of these problems can be naturally solved in
D--term inflation
in the framework of type I string theory rather than heterotic string theory.
The first problem is solved because in type I string theory the scale of the
anomalous D--term is not fixed; it is given by the VEV of a modulus which can
be of the required order of magnitude.
Moreover, unlike the heterotic string case in type I theory
there are gauge groups which can have rather small coupling constants. As a
result,
the coupling of $U(1)_A$ can be small enough to solve the second problem.

In the next section we briefly review the features type I strings (orientifolds
of  IIB string theory)
which are relevant for D--term inflation. In section 3, we show how these
features can be used to solve
the two problems that arise in heterotic string theories. We discuss our
results and conclude in the last section.

\bigskip
\centerline{\bf 2. Type I String Theory or Orientifolds of Type IIB Strings}
\medskip

Type I string compactifications with $N=1$ supersymmetry can be obtained by
orientifolds of type II strings[\AFIV]. We start with a type IIB string theory
in $D=10$ and mode it out by the world--sheet parity transformation $\Omega$.
This gives a type I string theory in $D=10$ with gauge group
$SO(32)$. The gauge group arises from the 32 D9 branes required for tadpole
cancellation.
This type I string theory is further compactified on an orbifold of $T^6$ (i.e.
on $T^6/\Gamma$ where $\Gamma$ is a discrete group such as $Z_n$)
resulting in a $D=4$ theory with $N=1$ supersymmetry and chiral matter
content. The above construction has only D9 branes but by considering more
elaborate orientifolds one can obtain models with two types of branes in the
theory. There can be two sets
of D--branes: either D9 and D5 branes or D3 and D7 branes[\AFIV]. The number of
each kind of brane is fixed again by tadpole cancellation. On the two different
kinds of branes ($Dp$ and $Dp^{\prime}$) there are gauge fields from strings
with both ends on the same kind of brane (i.e. $pp$ or $p^{\prime} p^{\prime}$
strings), giving two gauge groups, $G_p$ and $G_{p^\prime}$.
There are also matter fields which arise from strings with ends on different
kinds of branes (i.e. $pp^{\prime}$ strings).

For our purposes it is enough to consider the simplest case, a type IIB
orientifold with D3 branes and one set of D7 branes compactified on an orbifold
of $T^6$ with radius $R_c$. The D3 branes are along $X_{1,2,3}$ and the D7
branes are along $X_1, \ldots, X_7$. (Our results apply equally well
to the cases with more than one set of D7 branes or to the case with D9 and D5
branes.)
These models have two gauge groups; $G_3$ arising from D3 branes and $G_7$ from
D7
branes with couplings[\IMR]
$$\alpha_3={g_I \over 2} \qquad \alpha_7={g_I \over {2 M_I^4 R_c^4}} \eqno(9)$$
Here $g_I$ is the type I string coupling constant and $M_I=
\alpha_{str}^{-1/2}$ is the string scale.
Newton's constant is given by
$$G_N={1 \over M_P^2}= {g_I^2 \over {8M_I^8 R_c^8}} \eqno(10)$$
We see that contrary to the heterotic string case $M_P$ and $M_I$ do not need
to be of the same order of magnitude. From eqs. (9) and (10)  we get
$${\alpha_p M_P \over \sqrt 2}= {1 \over {M_I^{(p-7)} R_c^{(p-6)}}}=3.5 \times
10^{17}~ GeV \eqno(11)$$
assuming for one set of branes $\alpha_p \sim \alpha_U \sim1/25$, the unified
value of the Standard Model coupling constants. For the other set of branes we
get
$$\alpha_p={g_I \over 2} {1 \over (M_I R_c)^{(p-3)}} \eqno(12)$$
Note that there is freedom in the scales of $M_I$ and $R_c$ subject to eq.
(11). This
should be compared to the heterotic case for which $M_h= \sqrt{\alpha_U/8} M_P$
is fixed to be close to $M_P$ and independent
of the compactification radii.

The other important feature of type I string theory is related to the anomalous
D--term. In these
models the scale of the anomalous D--term is not fixed (compare to the
heterotic case with
eq. (2)) but is given by the VEV of some twisted moduli with a coefficient of
$O(1)$
[\IRU].\foot{Here the situation is different from the
one considered in [\JMR] since there are D branes in the vacuum.}
These moduli VEVs are related to the blowing up of the orbifold which smooths
out the singularities of the compact space. They are fixed only after
supersymmetry is broken but it is safe
to assume that the same mechanism that fixes the radii of the compactification
torus also
fixes them. (Note that these radii are given by VEVs of  the untwisted moduli.)
Thus, we assume
that untwisted moduli  VEVs are of order $R_c^{-1}$.

The matter content of these models arises from strings stretched between
different branes, i.e. 33, 37, 73 and 77 strings. We denote these fields by
$M^{33},M^{37},M^{73},M^{77}$.
$M^{33}$ and $M^{77}$ are in the adjoint representation of the respective gauge
groups, $G_3$
and $G_7$, whereas $M^{37}$ and $M^{73}$ are in the bifundamental
representation. In realistic models the gauge groups will be broken down by
Wilson lines and there will be gauge singlets coming from
either $M^{33}$ or $M^{77}$ (or both). The tree level superpotential
generically contains the terms[\IMR]
$$W=g_3(M^{33} M^{73} M^{37})+g_7(M^{77} M^{37} M^{73}) \eqno(13)$$

Note that the Yukawa couplings are given by the two gauge couplings in eq. (9).

\bigskip
\centerline{\bf 3. D--term Inflation in Type I String Theory}
\medskip

We now consider a type I string model such as the one above. This model has a
gauge singlet
$M^{33}$ (after symmetry breaking by Wilson lines) which we identify with the
inflaton field
$\sigma$. $M^{33}$ has tree level couplings to other gauge nonsinglets such as
$M^{37},M^{73}$ given by eq. (13). These play the role of the trigger fields
$\phi,\bar \phi$.
Note that they are charged under $G_7$ so we assume that the anomalous $U(1)_A$
comes from this sector. Also $M^{37}$ and $M^{73}$ are conjugates so their
$U(1)_A$ charges
are opposite (which we take to be $\pm 1$). $g_3 \sim 1/2$ gives the Yukawa
coupling, i.e. $\lambda$
in eq. (1). Thus, this simple model has all the ingredients for D--term
inflation
such as the inflaton and trigger fields, an anomalous D--term and the correct
superpotential.

As mentioned above, in type I models the scale of the anomalous D--term is not
fixed but
given by the VEV of a twisted modulus $M_t$. This VEV should be of the order of
magnitude
of other moduli VEVs such as compactification radii.
On the other hand, we saw that in type I compactifications there is some
freedom in $M_I$ and $R_c$ subject to eq. (11). If we embed the Standard Model
inside the D3 branes (i.e. inside $G_3$) we get from eq. (12) $M_I \sim 2
\times10^{16} ~GeV$ and
$R_c^{-1} \sim 8 \times 10^{15} ~GeV$. Then, $M_t \sim R_c^{-1}$ is of the
correct order of magnitude to account for COBE data. This is the solution of
the first problem mentioned in the
introduction.

In addition, we saw that the gauge group arising from one set of D branes
(in our case $G_7$) can have a  relatively small coupling given by eq. (9).
With the above values for $M_I$ and $R_c$ we get for the coupling of the $U(1)$
(which comes from the D7 branes or $G_7$) $\alpha_7 \sim \alpha \sim 5 \times
10^{-4}$ which is a rather small value. Substituting this into eq. (7) we find
that the initial value for the inflaton should be
$\sigma_i \sim 0.1 M_P$. This is still much larger than the critical value
$\sigma_{cr}$ but small
enough for the supergravity approximation to string theory to be valid. In this
case, the final value of the inflaton is close to the critical value, $\sigma_f
\sim \sigma_{cr}$.
More complicated models with more than two sets of D--branes and unisotropic
tori can give
smaller $\alpha$ and therefore smaller $\sigma_i$. In these cases, if $U(1)_A$
arises from a set of D7 branes with two large (i.e. larger than $M_I^{-1}$)
dimensions from eq. (9) we see that
$\alpha$ can be quite small resulting in $\sigma_i$ as small as $\sigma_{cr}$.

\bigskip
\centerline{\bf 4. Discussion and Conclusions}
\medskip

In this letter, we showed that the two problems which are generic to D--term
inflation in heterotic
string models are absent in type I string models.  The first problem related to
the magnitude of the density fluctuations is solved by the low scale of the
anomalous D--term in these models.
The string scale and the compactification radii of type I strings are not fixed
and may be much smaller than those of the heterotic ones. The scale of the
anomalous D--term is of the same order of magnitude which is about the scale
needed to accomodate the COBE data.  The second problem related to the very
large field values of the inflaton is also
absent due to the very small gauge coupling of $U(1)_A$. This is possible in
type I models since
the gauge group arises from two types of D branes independently. One set of
branes gives
the Standard Model group (with $\alpha_U \sim 1/25$)
whereas $U(1)_A$ can arise from the other set of D branes. In this case, for
$R_C^{-1}>M_I$,
the $U(1)_A$ gauge coupling can be much smaller than $\alpha_U $. Then from eq.
(7) we find that the initial value of the inflaton is at most
$M_P/10$ which is small enough for the supergravity approximation to string
theory to be valid.

In this letter, we considered the simplest possible type I string model with
two sets of D branes on an isotropic $T^6$. This
can be easily generalized to more complicated models with four sets of D branes
(one set of D3 and three sets of D7 branes or D9 and D5 branes) and a torus
with different compactification radii. All of our results will also hold in
these cases.  However, due to the extra fields and gauge symmetries present
in these cases some other requirements such as reheating may be more easily
met.  We also mentioned that a smaller initial value for the inflaton is
possible in these cases.
In discussing the D--term inflation scenario above we made a few generic
assumptions such as the presence of gauge singlet fields and the properties of
$U(1)_A$.
It would be interesting to build realistic $D=4$ type I string models and see
if these are in fact
realized.
We think it is quite encouraging to find that the generic problems of D--term
inflation in heterotic string theory are easily solved in type I string models.

\bigskip
\centerline{\bf  Acknowledgements}
\medskip

We would like to thank Ignatios Antoniadis for drawing my attention to refs.
[11-13] and
Andrei Linde for reading the manuscript.

\refout
\end